%% file: qic2010.tex
\documentclass[twoside]{article}

\usepackage{qic,epsfig}
\usepackage{amsmath}
\usepackage{cite}

\begin{document}

\setlength{\textheight}{8.0truein}    

\runninghead{Information Reconciliation for QKD}{D. Elkouss, J. Martinez-Mateo and V. Martin}

\normalsize\textlineskip
\thispagestyle{empty}
\setcounter{page}{1}

\copyrightheading{0}{0}{2010}{000--000}

\vspace*{0.88truein}

\alphfootnote

\fpage{1}

\centerline{\bf INFORMATION RECONCILIATION}
\vspace*{0.035truein}
\centerline{\bf FOR QUANTUM KEY DISTRIBUTION}
\vspace*{0.37truein}
\centerline{\footnotesize DAVID ELKOUSS, JESUS MARTINEZ-MATEO, VICENTE MARTIN\footnote{vicente@fi.upm.es}}
\vspace*{0.015truein}
\centerline{\footnotesize\it Research group on Quantum Information and Computation\footnote{http://gcc.ls.fi.upm.es}}
\baselineskip=10pt
\centerline{\footnotesize\it Facultad de Inform\'atica, Universidad Polit\'ecnica de Madrid}
\baselineskip=10pt
\centerline{\footnotesize\it Campus de Montegancedo, 28660 Boadilla del Monte, Madrid, Spain}
\vspace*{0.225truein}
\publisher{(received date)}{(revised date)}

\vspace*{0.21truein}

\abstract{Quantum key distribution (QKD) relies on quantum and classical procedures in order to achieve the growing of a secret random string ---the key--- known only to the two parties executing the protocol. Limited intrinsic efficiency of the protocol, imperfect devices and eavesdropping produce errors and information leakage from which the set  of measured signals ---the raw key--- must be stripped in order to distill a final, information theoretically secure, key. The key distillation process is a classical one in which basis reconciliation, error correction and privacy amplification protocols are applied to the raw key. This cleaning process is known as information reconciliation and must be done in a fast and efficient way to avoid cramping the performance of the QKD system. Brassard and Salvail proposed a very simple and elegant protocol to reconcile keys in the secret-key agreement context, known as \textit{Cascade}, that has become the de-facto standard for all QKD practical implementations. However, it is highly interactive, requiring many communications between the legitimate parties and its efficiency is not optimal, imposing an early limit to the maximum tolerable error rate. In this paper we describe a low-density parity-check reconciliation protocol that improves significantly on these problems. The protocol exhibits better efficiency and limits the number of uses of the communications channel. It is also able to adapt to different error rates while remaining efficient, thus reaching longer distances or higher secure key rate for a given QKD system.}

\vspace*{10pt}

\vspace*{3pt}
\communicate{to be filled by the Editorial}

\vspace*{1pt}\textlineskip    

\input{introduction}

\input{reconciliation}

\input{cascade}

\input{rate-compatible}

\input{protocol}

\input{results}

\input{conclusions}

\input{acknowledgment}

\input{references}

\end{document}

%% file: introduction.tex
\section{Introduction}
\label{sec:introduction}

\noindent A quantum key distribution protocol is composed of two parts: a quantum and a classical one~\cite{Gisin_02}. The quantum part involves the actual transmission of qubits, its manipulation and detection, and it is performed using a quantum channel. The classical part is done through a public, albeit integrity-preserving, classical channel and involves basis reconciliation, error correction and privacy amplification protocols. The quantum part results in the production of a raw key at both ends of the quantum channel. The raw key must be cleaned from all the unavoidable errors produced in this part. These include the intrinsic ones, due to the limited efficiency of the protocol, and those arising either from the inevitable imperfections in the physical setup or from eavesdropping. Intrinsic errors are easier to correct and this is usually done by bookkeeping of the detection events and subsequent discussion over the classical channel about the preparation state of the qubits leading to the recorded events. 
For example, in the standard BB84 protocol~\cite{Bennett_84}, half of the qubits detected by Bob will be in a base orthogonal to the one in which they were prepared by Alice, leading to a 50\% of detections that do not directly contribute bits to the final secret key. In a real setup, the remaining bits will still be affected from errors arising either from the physical implementation itself or from an eavesdropper, these being in principle indistinguishable. The process to clean the key from the errors is known as information reconciliation and is done through the classical channel.

The existence of optimal, although inefficient, protocols leaking a minimum of information in the process was demonstrated in~\cite{Brassard_94}. There, a practical protocol trading an acceptable amount of leaked information for efficiency was also proposed. This reconciliation protocol, known as \textit{Cascade}, has become the de-facto standard for all QKD practical implementations. However, it has several shortcomings that make it less than ideal under certain situations that are expected to become more common in real world environments.  Work has been done to improve on \textit{Cascade}~\cite{Sugimoto_00, Liu_03, Buttler_03, Han_09}, but none of the resulting methods have become as widespread.

Recent advances in QKD systems have seen a tremendous increase in key generation speed~\cite{Shields_08, Stucki_05, Gordon_05}. Current generation systems can be successfully used over longer distances or in noisier environments than before, like those arising when integration with conventional networks is required~\cite{Lancho_09, Zbinden_09}. This changes indicate that the reconciliation protocol must be efficient at high key and error rates. 

\textit{Cascade} is a highly interactive protocol that requires a high number of uses of the public channel to proceed. The number of uses raises markedly with the quantum bit error rate (QBER), thus it is not well suited for next generation QKD systems. From a practical point of view, the protocol must be implemented using two computers at both ends of the quantum channel that process the key and communicate through the public channel. 
The typical access latencies to the network are much higher than CPU operations, hence it is easy to produce a bottleneck  in highly interactive protocols. In fact, if a specialised communications network is not used, the communications needs of \textit{Cascade} are already the limiting factor in many situations and with current systems, instead of the much more delicate quantum part. On the other hand, a less than ideal efficiency limits  the maximum number of errors that can be corrected without publishing too much information, thus reducing the performance of the system when working at a high error rate.  

In this paper we describe a reconciliation protocol that overcomes these problems. The protocol exhibits a better efficiency than \textit{Cascade}, it can be adapted to a varying QBER with a low information leakage, extending the usable range of the system. It limits the number of uses of the public channel and its structure allows for a hardware implementation, avoiding communications and CPU bottlenecks, thus being well suited to next generation QKD systems in demanding environments. LDPC codes also have an structure that make them well suited for hardware implementations 

The paper is organised as follows: In Section~\ref{sec:reconciliation}, the information reconciliation problem in the secret-key agreement context is described and the current status of error correction in QKD is discussed. A new Information Reconciliation Protocol able to adapt to different channel parameters is presented and its asymptotic behavior discussed in Section~\ref{sec:rate-compatible}. In Section~\ref{sec:simulation-results} the results of a practical implementation of the protocol are shown. In particular the efficiency of the protocol is compared to its optimal theoretical value and to \textit{Cascade}.

%% file: reconciliation.tex
\section{Problem statement}
\label{sec:reconciliation}

\subsection{Information reconciliation}

Let Alice and Bob be two parties with access to dependent sources identified by two random variables, $X$ and $Y$ respectively. Information reconciliation is the process by which Alice and Bob extract common information from their correlated sources. In a practical setting Alice and Bob hold $\mathbf{x}$ and $\mathbf{y}$, two $n$-length strings that are the outcome of $X$ and $Y$, and they will agree in some string $\mathbf{s} = f(\mathbf{x}, \mathbf{y})$ through one-way or bidirectional conversation~\cite{VanAssche_06}. The conversation $\phi(\mathbf{x}, \mathbf{y})$ is also a function of the outcome strings, and its quality can be measured by the number of symbols involved in the conversation $M = |\phi(\mathbf{x}, \mathbf{y})|$.

\begin{figure} [htbp]
\centerline{\epsfig{file=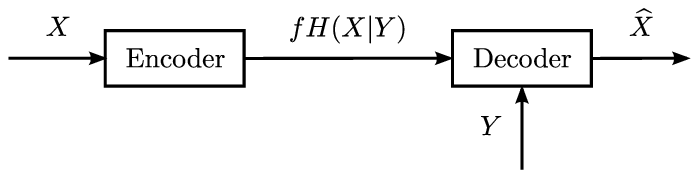, width=9cm}} 
\vspace*{13pt}
\fcaption{\label{fig:side-information}Source coding with side information.}
\end{figure}

Now, the problem of encoding correlated sources is a well known problem in information theory. To independently encode $X$ and $Y$ at least a rate $R \geq H(X) + H(Y)$ is needed. However, in their seminal paper, Slepian and Wolf~\cite{Slepian_73} demonstrated that to jointly encode both variables it is enough with a rate $R \geq H(X, Y)$ even if $X$ and $Y$ are encoded separately. Moreover, if $Y$ is available at the decoder only a rate of  $R\geq H(X|Y)$ is needed to encode $X$ (see Fig.~\ref{fig:side-information}), which in the information reconciliation context amounts for the minimum information needed in order to reconcile Alice's and Bob's strings. To measure the quality of a real reconciliation schema, that in a practical setting will encode $X$ with a higher rate than $H(X|Y)$, we use the efficiency parameter $f \geq 1$ defined as:

\begin{equation}
I_\textrm{real} = f H(X|Y) \geq I_\textrm{opt}
\label{eq:efficiency}
\end{equation}

\noindent where $I_\textrm{real}$ is the information published during the reconciliation process, $I_\textrm{opt}$ is the minimum information that would allow to reconcile Alice's and Bob's strings, and $H(X|Y)$ is the conditional Shannon entropy.

However, there are two other parameters to consider when evaluating the quality of a information reconciliation procedure: that is the computational complexity and the interactivity. The first one stresses that a real information reconciliation procedure must be feasible. Any sufficiently long random linear code of the appropriate rate could solve the problem~\cite{Zamir_02}, however optimal decoding is in general an NP-hard problem. The interactivity of a reconciliation protocol has to be taken into account because, specially in high latency scenarios, the communications overhead can pose a severe burden on the performance of the QKD protocol.

In order to evaluate the quality of the reconciliation, we will concentrate in discrete variable QKD protocols even if the ideas presented here can be easily extrapolated to other scenarios. Most QKD protocols encode the information in discrete variables~\cite{Bennett_84, Bennett_92}, although there are many proposals on continuous variable protocols \cite{Ralph_99, Grosshans_02, Fossier_09}. Errors on the quantum channel are normally uncorrelated and symmetric or, if prior to the reconciliation Alice and Bob apply a random permutation, they can behave as such~\cite{Gottesman_03}. In this situation Alice's and Bob's strings can be regarded as the input and output of a binary symmetric channel (BSC), characterized by the crossover probability $\varepsilon$, and the efficiency parameter $f$ can be described as the relationship between the length of the conversation $M = |\phi(\mathbf{x}, \mathbf{y})|$ and the optimal value $N\cdot H(X|Y) = N\cdot h(\varepsilon)$:

\begin{equation}
f = \frac{M}{N\cdot h(\varepsilon)}
\label{eq:f}
\end{equation}

It was first shown in~\cite{Liveris_02} that low-density parity-check (LDPC) codes used within Wyner's coset scheme~\cite{Wyner_74, Zamir_02} are a good solution for the compression of binary sources with side information. LDPC codes~\cite{Richardson_01a} are linear codes that have a sparse parity check matrix. These codes, when decoded with the belief propagation algorithm, can perform very close to the theoretical limit.

The fundamental idea is to assign each source vector to a bin from a set of $2^{H(X|Y) + \iota}$ known bins, the encoder describes the bin to the decoder, and the decoder searches for the source vector inside the described bin. Let $\mathbf{x}$ and $\mathbf{y}$ be two binary strings of length $n$, and $C$ a $[n,k]$  binary linear code specified by its parity matrix $H$. The syndrome $\mathbf{s}$ of a vector $\mathbf{x}$ is the $n-k$ string defined as $\mathbf{s} = H \mathbf{x}$ being $\mathbf{s} = \mathbf{0}$ for all the codewords in $C$. Each syndrome $\mathbf{s}$ defines a coset $C_\mathbf{s}$ as the set of all strings, $\{ \mathbf{x} \}$, that verify $H \mathbf{x} = \mathbf{s}$. Wyner's schema consists in assigning each source vector to one of the cosets of $C$. Encoding $\mathbf{x}$ amounts then to compute its syndrome, and decoding is simply to find the member in $C_\mathbf{s}$ closest to $\mathbf{y}$. An LDPC message passing decoder was modified in \cite{Liveris_02} to take into account the syndrome decoding proposed by Wyner. This same procedure can be applied to the QKD scenario.

%% file: cascade.tex
\subsection{Previous work}
\label{sec:cascade}




As mentioned in the introduction, the most widely used and best known protocol for error correction in the QKD context is \textit{Cascade}. Proposed by Brassard and Salvail~\cite{Brassard_94}, this protocol runs for a fixed number of passes. In each pass, Alice and Bob divide their strings into blocks of equal length. The initial block length depends on the estimated error probability, $p$, and it is doubled when starting a new pass. For each block they compute and exchange its parity. A parity mismatch implies an odd number of errors, and a dichotomic search allows both parties to find one of the errors. Whenever an error is found after the first pass, it uncovers an odd number of errors masked on the preceding passes and the algorithm returns to correct those errors previously undetected. This cascading process gives name to the protocol. Several papers propose improvements on \textit{Cascade}~\cite{Sugimoto_00, Liu_03}, these papers analyse how the block length is to be chosen and increased in order to optimise the efficiency of the reconciliation, but the main characteristics remain unaltered. \textit{Cascade}, although highly interactive, is reasonably efficient and easy to implement. It is well known and has become the de-facto standard, hence we have chosen \textit{Cascade} as the benchmark to compare against.

\textit{Winnow}~\cite{Buttler_03} is another well know reconciliation protocol in the QKD context, it requires only two communications between the parties. In the first communication Alice and Bob exchange the parities of every block. After that, they exchange the syndrome of a Hamming code for correcting single errors in each block with a parity mismatch. The protocol incorporates a privacy maintenance procedure by discarding one bit per parity revealed (i.e. $m$ bits are discarded when a syndrome of length $m$ is exchanged). Its main advantage is a reduction on the number of communication needed, however the efficiency of the protocol is worse than that of \textit{Cascade} in the error range of interest. Recently, some interesting improvements have been proposed for selecting an optimum block length in this protocol~\cite{Han_09}. 

Modern coding techniques have not been applied to discrete variable QKD until recently. LDPC codes were proposed and used on~\cite{Elliott_05}. But as the codes had not been specifically designed for the problem, aside from the inherent advantage of forward error correction, the efficiency was worse than that of \textit{Cascade}. These codes have been also used in the context of continuous variable QKD~\cite{Fossier_09}. LDPC codes were first optimized for the BSC on \cite{Elkouss_09}, and although the results were close to optimal for the designed codes, the efficiency curve exhibited a saw behaviour due to a lack of information rate adaptability in the proposed procedure~\cite{Elkouss_10}. Since the error rate can vary among transmissions, it is important for a protocol to be able to cope with this change.

%% file: rate-compatible.tex
\section{Rate-compatible reconciliation}
\label{sec:rate-compatible}

\noindent Although linear codes are a good solution for the reconciliation problem, since they can be tailored to a given error rate, their efficiency degrades when it is not known beforehand. This is the case in QKD, where the error rate is an a priori unknown that is estimated for every exchange. The QBER might vary significantly in two consecutive key exchanges, specially when the quantum channel is transported through a shared optical fibre that can be used together with several independent classical or quantum channels that can add noise. To address this problem there are two different options: (i) it is possible to build a code once the error rate has been estimated, and (ii) a pre-built code can be modified to adjust its information rate. The computational overhead would make the first option almost unfeasible except for very stable quantum channels, something difficult to achieve in practise and impossible in the case of a shared quantum channel in a reconfigurable network environment~\cite{Lancho_09}. In this paper we propose the use of the second strategy as the easiest and most effective way to obtain a code for the required rate, for which we describe a protocol that adapts pre-built codes in real time while maintaining an efficiency close to the optimal value.

\subsection{Rate modulation}
\label{sec:rate-modulation}

\begin{figure} [htbp]
\centerline{\epsfig{file=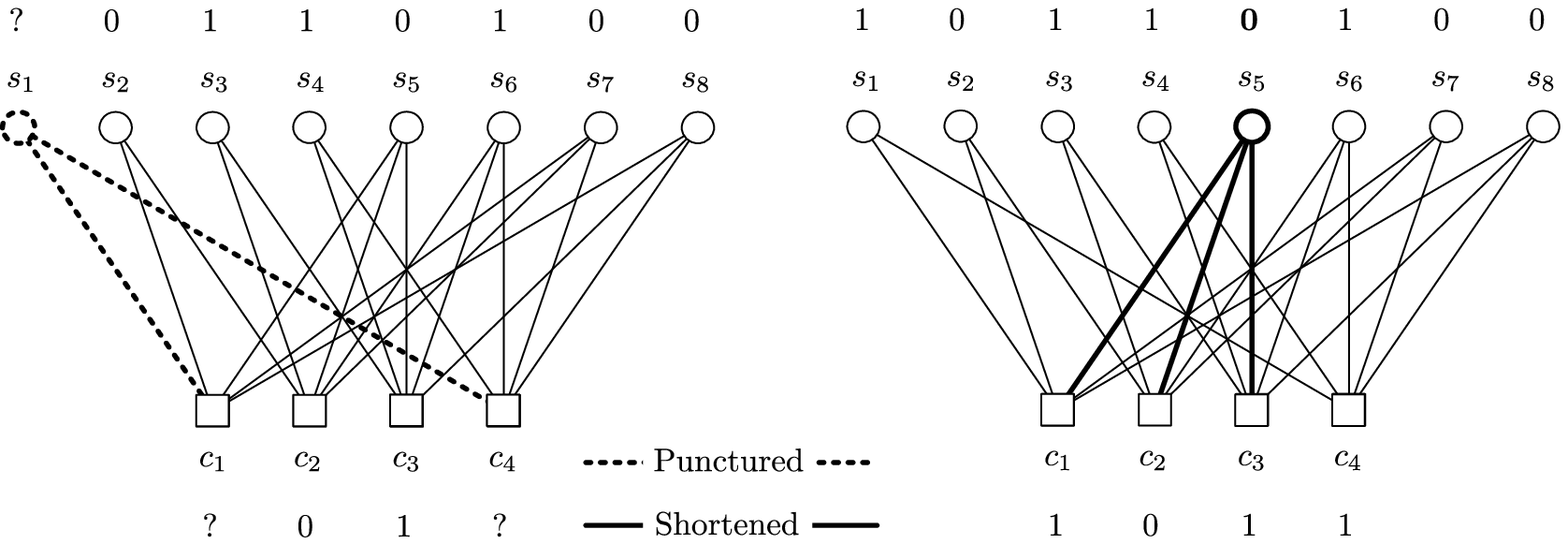, width=\textwidth}} 
\vspace*{13pt}
\fcaption{Examples of puncturing and shortening strategies applied to a linear code represented by its Tanner graph. In the puncturing example (left) one symbol is deleted from the word and a [8,4] code, with rate $R=1/2$, is converted to a [7,4], increasing its rate to $R=4/7$. In the shortening example (right), one symbol is deleted from the encoding and the same [8,4] code is converted to a [7,3] code, the rate now decreases to $R=3/7$.}
\label{fig:puncturing-shortening}
\end{figure}

Puncturing and shortening are two common strategies used to adapt the rate of a linear code. This process of adapting the information rate of a pre-built code will be referred as rate modulation. When $p$ punctured symbols of a codeword are removed, a $[n, k]$ code is converted into a $[n-p, k]$ code. Whereas, when shortening, $s$ symbols are removed during the encoding process, and a $[n, k]$ code is converted into a $[n-s, k-s]$ code. A graphical representation, on a Tanner graph, of the procedures just described for puncturing and shortening and its effects on the rate of the sample code is shown in Fig.~\ref{fig:puncturing-shortening}.

\begin{figure} [htbp]
\vspace*{13pt}
\centerline{\epsfig{file=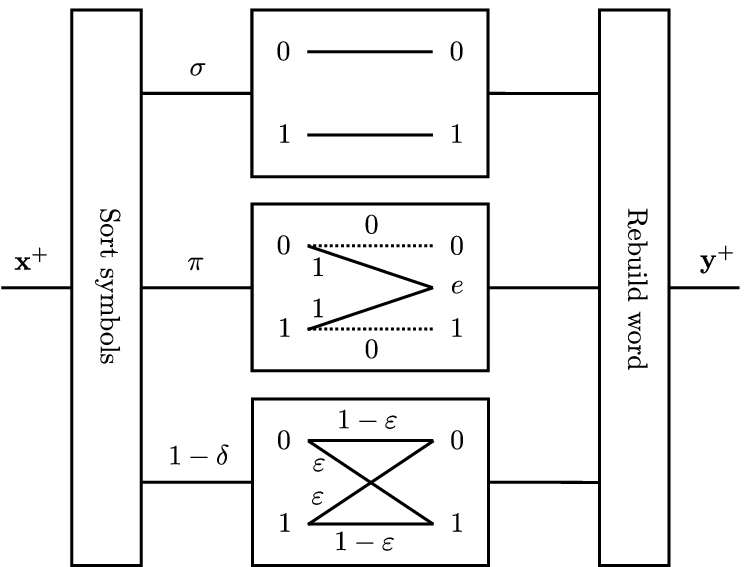, width=8cm}} 
\vspace*{13pt}
\fcaption{\label{fig:channel-model}Channel model. Puncturing and shortening on a LDPC code results in the division of the original binary symmetric channel used to reconcile Alice's {$\mathbf x$} string with Bob's {$\mathbf y$}  into three different channels: a binary erasure  channel with erasure probability of 1 (for the fraction $\pi$ of punctured symbols), a BEC with erasure probability of 0 (for the fraction $\sigma$ of shortened symbols) and a binary symmetric channel with crossover probability $\varepsilon$ (for the rest of the symbols). }
\end{figure}

These procedures may be regarded as the transmission of different parts of the codeword over different channels (see Fig.~\ref{fig:channel-model}). Since puncturing is a process by which $p$ codeword symbols are eliminated, it can be seen as a transmission over a binary erasure channel (BEC) with erasure probability of $1$, $\textrm{BEC}(1)$. Shortening is a process by which $s$ codeword symbols are known with absolute certainty, as such it can be seen as a transmission over a BEC with erasure probability of $0$, $\textrm{BEC}(0)$. The remaining symbols are transmitted by the real channel which in the present paper can be modelled by a binary symmetric channel with crossover probability $\varepsilon$, $\textrm{BSC}(\varepsilon)$

Supposing that $R_0$ is the original coding rate, the modulated rate is then calculated as:

\begin{equation}
R = \frac{R_0 - \sigma}{1 - \pi - \sigma} = \frac{k - s}{n - p - s}
\label{eq:rate}
\end{equation}

\noindent where $\pi$ and $\sigma$ represent the ratios of information punctured and shortened respectively.

Both strategies, puncturing and shortening, can be applied simultaneously. Given a $[n, k]$ code and $n'\leq n$ bits, if puncturing and shortening are applied with a constant number $d$ of punctured and shortened symbols, a single code can be used to protect the $n'$ bits for different error rates. There are two consequences of applying a constant $d$: (i) there is a limit to the minimum and maximum achievable information rates. These limits, expressed as a function of $\delta = d  / n$, define the correction interval:

\begin{equation}
0 \leq R_{\textrm{min}} = \frac{R_0 - \delta}{1 - \delta} \leq R \leq \frac{R_0}{1 - \delta} = R_{\textrm{max}} \leq 1
\label{eq:rate-interval}
\end{equation}

\noindent (ii) puncturing and shortening procedures cause an efficiency loss~\cite{Ha_04}. Therefore, there is a tradeoff between the achievable information rates and reconciliation efficiency. 

\begin{figure} [htbp]
\vspace*{5pt}
\centerline{\epsfig{file=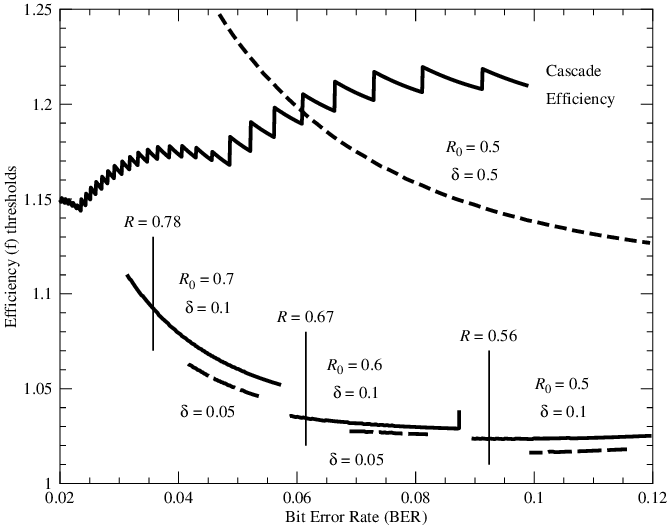}}
\vspace*{5pt} 
\fcaption{\label{fig:efficiency-thresholds}Efficiency thresholds for different codes with information rates, $R_0 = 0.5$, $0.6$ and $0.7$ as a function of the quantum bit error rate (QBER). Two $\delta$ values, 0.1 (solid line) and 0.05 (dashed) have been used to adapt the rate for each code. As a comparison, a single code covering all of the QBER range of interest, with rate $R_0= 0.5$ and  $\delta=0.5$, is presented to show how the efficiency degrades for high  $\delta$ values, although a broader range is covered. The codes have been optimised using the density evolution algorithm for the BSC. The \textit{Cascade} efficiency was calculated using the same sample size ($2 \times 10^5$). The block size used in the first step, $k_1$, is given by $k_1 = \lceil 0.73/QBER \rceil$ (optimized in~\cite{Crepeau_95}) and doubled in every subsequent step $k_n = 2 k_{n-1}$. The sawtooth behaviour of the \textit{Cascade} efficiency reflects the points where $k_1$ changes.} 
\end{figure}

This efficiency loss, caused by high levels of puncturing and shortening, can be avoided if a set of $n$ codes $\zeta_i$ with different information rates is used: $R_0(\zeta_1)\le R_0(\zeta_2)\le R_0(\zeta_n)$. The target error range can then be partitioned into, $[R_{\textrm{min}}(\zeta_1), R_{\textrm{max}}(\zeta_1)] \cup [R_{\textrm{min}}(\zeta_2), R_{\textrm{max}}(\zeta_2)]\cup ... \cup [R_{\textrm{min}}(\zeta_n), R_{\textrm{max}}(\zeta_n)]$, not necessarily with the same size. The number of intervals depends on the width of the error rate range to cover and on the desired efficiency. The compromise between the width of the interval covered and the achieved efficiency in the one code case is transferred to a compromise between efficiency and the added complexity of managing several codes. Fig.~\ref{fig:efficiency-thresholds} shows the computed efficiency thresholds for several families of codes with different coding rates.

%% file: protocol.tex
\subsection{Protocol}
\label{sec:protocol}

We now proceed to describe a rate-compatible information reconciliation protocol using puncturing and shortening techniques as described above.

{\it Step 0: Raw key exchange}. 
Alice and Bob obtain a 
raw key by running a QKD protocol through a quantum channel (see Section~\ref{sec:reconciliation}). This key exchange may be modelled as follows. Alice sends to Bob the string $\mathbf{x}$, an instance of a random variable $X$, of length $\ell = n - d$ through a binary symmetric channel with crossover probability $\varepsilon$, BSC($\varepsilon$) (or a black box behaving as such). Bob receives the correlated string, $\mathbf{y}$, but with discrepancies to be removed in the following steps.

{\it Step 1: Pre-conditions}. Prior to the key reconciliation process Alice and Bob agree on the following parameters: (i) a pool of shared codes of length $n$, constructed for different coding rates; (ii) the size of the sample, $t$, that will be used to estimate the error rate in the communication; and (iii) the maximum number of symbols that will be punctured or shortened to adapt the coding rate, $d = p + s = n\delta$.

{\it Step 2: Error rate estimation}. Bob chooses randomly a sample of $t$ bits of $\mathbf{y}$, $\alpha(\mathbf{y})$, and sends them and their positions, $\beta(\mathbf{y})$, to Alice through a noiseless channel (i.e. the public and integrity-preserving channel used in the classic part of a QKD protocol). Using the positions received from Bob, $\beta(\mathbf{y})$, Alice extracts an equivalent sample in $\mathbf{x}$, $\alpha(\mathbf{x})$, and estimates the crossover probability for the exchanged key by comparing the two samples:

\begin{equation}
\varepsilon' = \frac{\alpha(\mathbf{x}) + \alpha(\mathbf{y})}{t}
\end{equation}

Once Alice has estimated $\varepsilon'$, she knows the theoretical rate for a punctured and shortened code able to correct the string. Now she computes the optimal rate corresponding to the efficiency of the code she is using: $R = 1 - f(\varepsilon') h(\varepsilon')$; where $h$ is the binary Shannon entropy function and $f$ the efficiency. Then she can derive the optimal values for puncturing and shortening, $p$ and $s$ respectively, as:

\begin{equation}
\begin{split}
s & = \lceil (R_0 - R (1 - d / n)) \cdot n \rceil \\
p & = d - s
\end{split}
\end{equation}

{\it Step 3: Coding}. Alice creates a string $\mathbf{x^+} = g(\mathbf{x}, \mathbf{\sigma_{\varepsilon'}}, \mathbf{\pi_{\varepsilon'}})$ of size $n$. The function $g$ defines the $n - d$ positions to take the values of string $\mathbf{x}$, the $p$ positions to be assigned random values, and the $s$ positions to have values known by Alice and Bob. The set of $n - d$ positions, the set of $p$ positions and the set of $s$ positions and their values come from a synchronised pseudo-random generator. She then sends $s(\mathbf{x^+})$, the syndrome of $\mathbf{x^+}$, to Bob as well as the estimated crossover probability $\varepsilon'$.

This process can be regarded as jointly coding (and decoding) the original strings sent through a BSC(QBER) with $p$ bits sent through a binary erasure channel (BEC) with erasure probability 1, and $s$ bits sent through a noiseless channel (see Fig.~\ref{fig:channel-model}).

{\it Step 4: Decoding}. Bob can reproduce Alice's estimation of the optimal rate $R$, the positions of the $p$ punctured bits, and the positions and values of the $s$ shortened bits. Bob then creates the corresponding string $\mathbf{y^+} = g(\mathbf{y}, \mathbf{\sigma_{\varepsilon'}}, \mathbf{\pi_{\varepsilon'}})$. He should now be able to decode Alice's codeword with high probability, as the rate has been adapted to the channel crossover probability. Bob sends an acknowledgement to Alice to indicate if he successfully recovered $\mathbf{x^+}$.

\begin{figure} [htbp]
\centerline{\epsfig{file=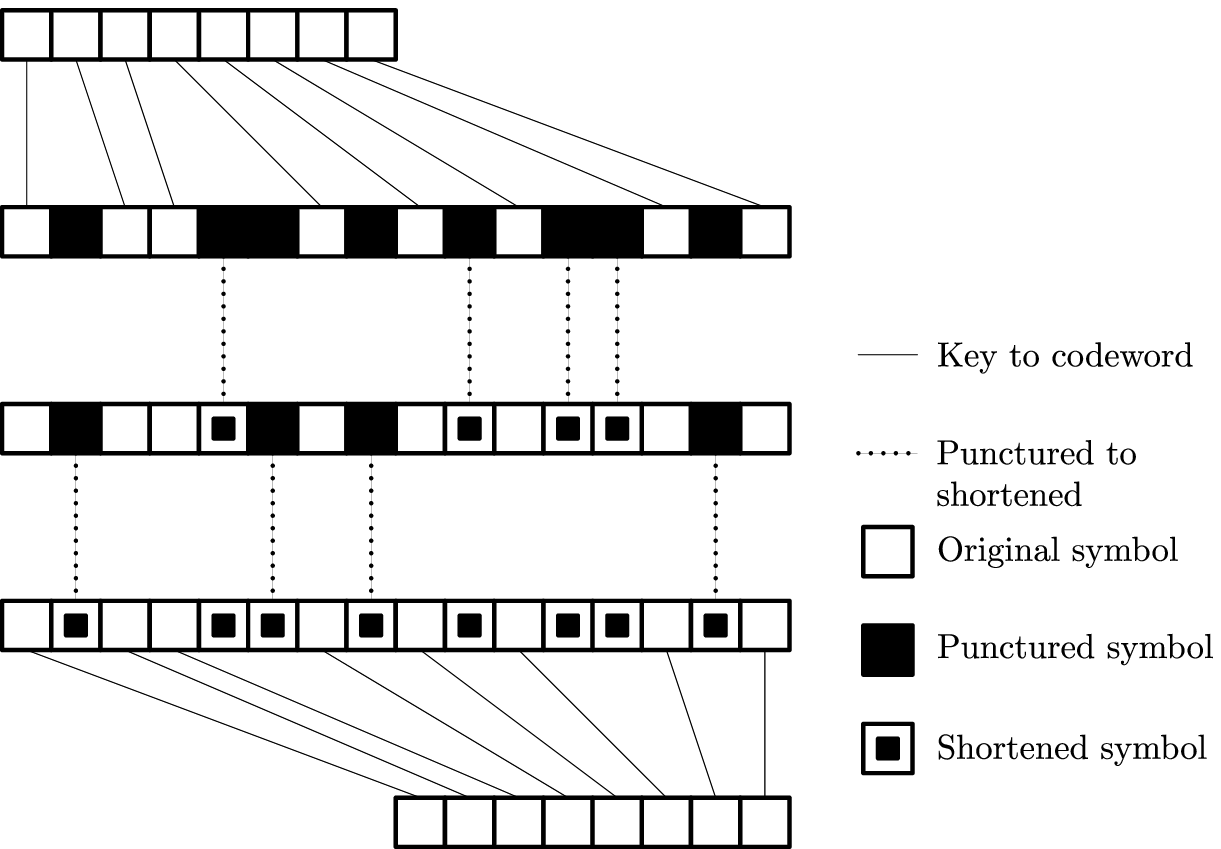, width=9cm}} 
\vspace*{13pt}
\fcaption{\label{fig:interactive-protocol}Protocol sequence for interactive reconciliation. The figure shows how punctured symbols are converted into shortened symbols in each round, thus repeating coding and decoding steps for different coding rates. The interactive protocol concludes when the target string has been reconciled or there are no more punctured symbols to be revealed.}
\end{figure}

{\it Step 5: (Optional) Interactive decoding}. If Bob does not succeed in recovering Alice's string, Alice can reduce the information rate of the code by revealing some $r^* \le p$ of the punctured bits on the public channel, that become shortened bits (see Fig.~\ref{fig:interactive-protocol}). Steps 3 and 4 are then repeated, Alice computes the new syndrome and sends it to Bob, who tries to decode and send an acknowledge to Alice. Let $p^{(i)}$ and  $s^{(i)}$ be the number of punctured and shortened bits respectively in the $i$-th round of the proposed protocol, and $r^{(i+1)}$ the number of punctured bits to be revealed for the next round, the new proportion of punctured and shortened bits used for the reconciliation are calculated as $p^{(i+1)} = p^{(i)} - r^{(i+1)}$ and $s^{(i+1)} = s^{(i)} + r^{(i+1)}$, respectively. These steps can be repeated while Bob does not find the correct string and there are punctured bits that have not been revealed as shortened bits, i.e. while $p \ge 0$.

%% file: results.tex
\section{Simulation results}
\label{sec:simulation-results}

In this section we discuss the efficiency of the rate-compatible information reconciliation protocol without the interactive decoding step, comparing the results of the protocol to regular LDPC codes as proposed in~\cite{Elkouss_09} and to~\textit{Cascade}. The purpose of this simulations is to highlight that the proposed protocol extends the working range in QKD, allowing to distill a key in a wider QBER range than previous information reconciliation protocols.

\begin{figure} [htbp]
\vspace*{13pt}
\centerline{\epsfig{file=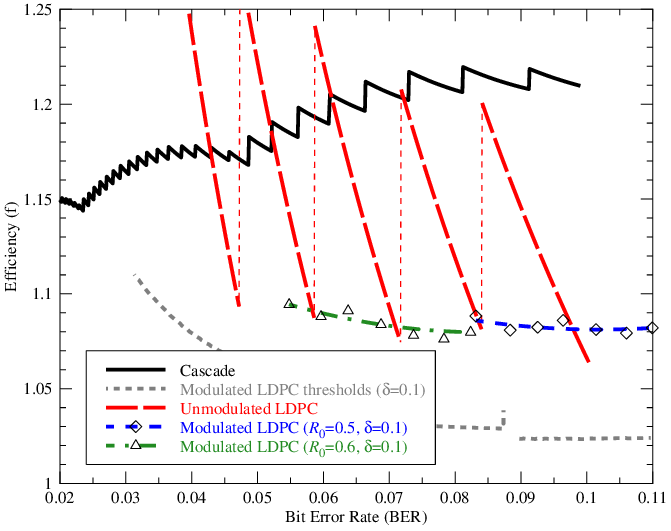}}
\vspace*{13pt}
\fcaption{\label{fig:simulations-results}Computed efficiency for medium to high error rates, a typical 
range expected in shared quantum channel environments, long distances or high losses scenarios, such as in networks, and where obtaining high efficiency is critical. The solid line is the \textit{Cascade} efficiency. Its parameters are the same than for Fig.~\ref{fig:efficiency-thresholds}. The dotted line represents the modulated LDPC thresholds. For all LDPC results shown here $\delta=0.1$. The long, thick, dashed lines joined by thin dashed lines is the efficiency of an unmodulated code. Short dash and dash-dotted lines are the results for the modulated codes. Dash-dotted is for a rate $R_0 = 0.6$ and short dash are for $R_0 = 0.5$, triangles and diamonds are used to mark the computed points. The smooth and efficient behaviour of the modulated, rate adapted codes, as compared to the unmodulated version is to be noted. The gain in efficiency over \textit{Cascade} allows for an extended usability range of the system at high QBER.}
\end{figure}

Fig.~\ref{fig:simulations-results} shows the efficiency, calculated as defined in Eq.~(\ref{eq:efficiency}), in the reconciliation process simulated for three different alternatives: (i) using the \textit{Cascade} protocol, (ii) using LDPC codes without adapting the information rate, and (iii) using LDPC codes adapting the information rate with the rate-compatible protocol proposed here. The target error range selected is $[0.055,0.11]$, where a high efficiency protocol is a must. Low QBER rates do not demand a close to optimal efficiency since other requisites, such as the throughput, are more critical in obtaining a high secret key rate.  In order to achieve a efficiency close to $1$, the error range $[0.055,0.11]$ has been divided into two correction intervals: $R_0(\zeta_1)=0.5$, $R_0(\zeta_2)=0.6$ and $\delta=0.1$. The codes have been constructed using families of LDPC codes specifically optimised for the BSC. Generating polynomials of these families can be found in~\cite{Elkouss_09}, however they were not designed for shortening and puncturing. Taking into account these parameters in the generating polynomial design process would allow to cover the whole QBER range with high efficiency.

The construction process has been also optimised using a modified progressive edge-growth algorithm for irregular codes with a detailed check node degree distribution~\cite{Martinez_10}. A codeword length of $2 \times 10^5$ bits has been used.

The results show that there is a small price to pay for the rate adaptation. LDPC codes without puncturing and shortening behave slightly better near their threshold, however for the $\delta$ value chosen the penalty is very small and the rate-compatible protocol allows to reconcile strings in all the range with $f \le 1.1$. The unmodulated LDPC codes exhibit an undesirable saw behaviour that can lead to efficiencies worse than that of \textit{Cascade} unless many different codes are calculated, incurring in an unacceptable penalty in CPU time. The new protocol works at a much better efficiency than \textit{Cascade}, that performs in all the tested range with $f \ge 1.17$.

%% file: conclusions.tex
\section{Conclusions}
\label{sec:conclusions}

We have demonstrated how to adapt an LDPC code for rate compatibility. The capability to adapt to different error rates while minimizing the amount of published information is an important feature for secret-key reconciliation in the QKD context, specially whenever it is used in long distance links or in noisy environments such as those arising in shared optical networks. In these demanding environments high efficiency is necessary to distill a key. The protocol improves on \textit{Cascade}, allowing to reach efficiencies close to one while limiting the information leakage and having the important practical advantage of low interactivity: only one message is exchanged by both parties. This high efficiency allows to extend the working range in QKD, that is, it allows to distill a key in a wider QBER range than previous information reconciliation protocols.


%% file: acknowledgment.tex
\section*{Acknowledgment}

\noindent This work has been partially supported by the project Quantum Information Technologies in Madrid\footnote{http://www.quitemad.org} (QUITEMAD), Project P2009/ESP-1594, \textit{Comunidad Aut\'onoma de Madrid}.

The authors gratefully acknowledge the computer resources, technical expertise and assistance provided by the \emph{Centro de Supercomputaci\'on y Visualizaci\'on de Madrid}\footnote{http://www.cesvima.upm.es} (CeSViMa) and the Spanish Supercomputing Network.

%% file: references.tex
\nonumsection{References}